\documentclass[useAMS,usenatbib]{mn2e}
\usepackage{times}
\usepackage{amsmath}
\usepackage{units}
\usepackage{hyperref}
\usepackage{color}

\newcommand{\kms}{{~\rm km\; s^{-1}}}
\newcommand{\cc}{{~\rm cm^{-3}}}

\newcommand{\s}{{~\rm s}}

\newcommand{\g}{{~\rm g}}
\newcommand{\K}{{~\rm K}}
\newcommand{\erg}{{~\rm erg}}
\newcommand{\yr}{{~\rm yr}}

\newcommand{\days}{{~\rm days}}
\newcommand{\Rsun}{{~\rm R_{\sun}}}
\newcommand{\Msun}{{~\rm M_{\sun}}}
\newcommand{\ergs}{{~\rm erg~s^{-1}}}

\newcommand{\aap}{A\&A}
\newcommand{\actaa}{Acta Astron.}
\newcommand{\apj}{ApJ}
\newcommand{\apjl}{ApJL}
\newcommand{\apjs}{ApJS}
\newcommand{\araa}{ARA\&A}
\newcommand{\mnras}{MNRAS}
\newcommand{\nat}{Nature}

\newcommand{\revision}[1]{#1}

\begin{document}

\title[Constraining the DD scenario for SNe~Ia from merger ejected matter]
{	Constraining the double-degenerate scenario for Type Ia supernovae from merger ejected matter}

\author[N. Levanon, N. Soker and E. Garc{\'{\i}}a--Berro]
{
	Naveh Levanon,$^{1,}$\thanks{Emails: \href{mailto:nlevanon@tx.technion.ac.il}{nlevanon@tx.technion.ac.il},
		\href{mailto:soker@physics.technion.ac.il}{soker@physics.technion.ac.il}}
 	Noam Soker$^{1,}$\footnotemark[1]
 	and Enrique Garc{\'{\i}}a--Berro$^{2,3}$ \\ 	
 	$^{1}$Department of Physics, Technion -- Israel Institute of Technology, Haifa 32000 Israel \\
 	$^{2}$Departament de F{\'{\i}}sica Aplicada, Universitat Polit\`ecnica de Catalunya, c/Esteve Terrades 5, 08860 Castelldefels, Spain \\
 	$^{3}$Institute for Space Studies of Catalonia, c/Gran Capit\`a 2-4, Edif. Nexus 201, 08034 Barcelona, Spain
}

\maketitle

\begin{abstract}

We follow the mass expelled during the WD-WD merger process in \revision{a particular case of}
the Double-Degenerate (DD) scenario for Type Ia supernovae (SNe~Ia), and find that
the interaction of the SN ejecta with the resulting wind affects the early (first day)
light curve in a way that \revision{may be in conflict with} some SN~Ia observations, if the detonation
occurs shortly after the merger (i.e., $10^3~{\rm sec} \la t_{\rm exp} \la 1~{\rm day}$).
The main source of the expelled mass is a disk-wind, or jets that are launched by the accretion
disk around the more massive WD during the viscous phase of the merger.
This disk-originated matter (DOM) will be shocked and heated by the SN ejecta from an explosion,
leading to additional radiation in the early lightcurve.
This enhanced early radiation could then be interpreted as an explosion originating from a
progenitor having an inferred radius of one solar radius or more, in conflict with
observations of SN~2011fe.

\end{abstract}


\section{Introduction}
\label{sec:introduction}

There is no consensus on the evolutionary routes that bring CO
white dwarfs (WDs) in binary systems to explode as Type Ia
supernovae (SNe~Ia; e.g., \citealt{Livio2001, Maoz2010, Howell2011,
Maozetal2014, Ruiz-Lapuente2014, TsebrenkoSoker2015b}). The
different scenarios currently considered
\revision{can be summarized as follows (for more details and a
table comparing the pros and cons of the five scenarios see
\citealt{TsebrenkoSoker2015b}).}
\newline (a) \emph{The double degenerate (DD) scenario} (e.g.,
\citealt{Webbink1984, Iben1984}).
In this scenario two WDs orbiting each other lose angular
momentum and energy through the radiation of gravitational
waves \citep{Tutukov1979}, and a merger occurs. The exact time
after merger when explosion occurs is unknown, and different
mechanisms are discussed in the literature (e.g.,
\citealt{vanKerkwijk2010}). This scenario can account for
sub-Chandrasekhar mass explosions as well (e.g.,
\citealt{vanKerkwijk2010, BadenesMaoz2012}). In recent years a
violent merger process was proposed as a channel to ignite the WD
(e.g., \citealt{Loren2010, Pakmor2013}). Others consider a very
long delay from merger to explosion, e.g., because rapid rotation
keeps the structure \revision{stable even as the accreting remnant exceeds
the Chandrasekhar mass
($M_{\rm Ch}$; \citealt{TornambePiersanti2013}).}
\newline (b) \emph{The core-degenerate (CD) scenario} \citep{Sparks1974,Livio2003,
KashiSoker2011, IlkovSoker2012, IlkovSoker2013, Soker2013,
Soker2014}. In this scenario a WD merges with a hot core of a
massive asymptotic giant branch (AGB) star. The explosion can
occur shortly after the common envelope (CE) phase, hence leading to a
SN~Ia inside a planetary nebula \citep{TsebrenkoSoker2013,
TsebrenkoSoker2015, TsebrenkoSoker2015b}, or after a very long time delay.
There is some overlap between the DD and CD scenarios,
\revision{in the sense that if the merger occurs after the
termination of the CE phase, but while the core is not yet on
the cooling track of a WD, both scenarios describe the same system.
Basically, this occurs when the merger occurs during the
planetary nebula phase of the system.}
\newline (c) \emph{The single degenerate (SD) scenario} (e.g.,
\citealt{Whelan1973, Nomoto1982, Han2004}). According to this
scenario a WD accretes mass from a non-degenerate stellar
companion, \revision{and explodes as it approaches the Chandrasekhar mass.}
There is also the scenario for accretion of helium-rich
material from a non-degenerate helium star (e.g.,
\citealt{Iben1987, Ruiteretal2011}), which we list under the
double-detonation scenario.
\newline (d) \emph{The double-detonation (DDet) mechanism}
(e.g., \citealt{Woosley1994, Livne1995}), in which a
sub-Chandrasekhar mass WD accumulates on its surface a layer of
helium-rich material, which can detonate and lead to a second
detonation near the center of the CO WD. One version has a helium
WD as the donor star (e.g., \citealt{Shenetal2013,Piersanti+2014}
for recent papers).
\revision{While \cite{Ruiteretal2011} found that a large fraction
of SN~Ia can be attributed to the sub-Chandrasekhar DDet scenario,}
\cite{Piersantietal2013} found that the DDet scenario can account
for only a small fraction of all SN~Ia. 
\revision{More recently \cite{Ruiteretal2014} argued that the DDet
scenario can account for a large fraction of SN~Ia if most ($>70 \%$)
of the donors are He~WDs.}
\cite{Papishetal2015} found that the explosion in the case of a
He~WD donor leads to a non-spherical SN remnant (SNR), 
\revision{and in the case of a close helium WD, the latter will be
ignited and will eject too much helium to be compatible with observations. 
\cite{Papishetal2015} also pointed out that the expected SNR has a dipole
asymmetry that no well-resolved SNR~Ia has.
Another issue with this scenario is that because the exploding CO~WD
barely grows before it explodes \citep{ShenBildsten2009}, the DDet
scenario predicts that most exploding WDs will have a mass of $<1.2 \Msun$.
This is at odds with recent findings that a large fraction of SN~Ia
masses are peaked around $1.4 \Msun$ \citep{Scalzoetal2014}.
\cite{Seitenzahletal2013} also claim that at least $50 \%$ of all SN~Ia
come from near-Chandrasekhar mass WDs.
It seems that the DDet scenario can lead to explosions similar to SN~Ia,
but not to common SN~Ia.}
\newline (e) \emph{The WD-WD collision scenario.} In this
scenario two WDs collide directly, either because of interaction
with a tertiary star (e.g., \citealt{Thompson2011, KatzDong2012,
Kushniretal2013}), or by random encounters in dense stellar
systems such as globular clusters (e.g.,
\citealt{Raskin+2009,Rosswog+2009,Aznar2014}). The collision sets
an immediate explosion. Despite some attractive features of this
scenario, it can account for at most few per cent of all SNe Ia
\citep{Hamersetal2013, Prodanetal2013, Soker2014}.

Each of these scenarios has some problems, and in some cases
these are severe \citep{Soker2014}.
Our view, based on the comparison tables of \cite{Soker2014} and
\cite{TsebrenkoSoker2015b}, is that the most promising are
the DD and CD scenarios.
With this view in mind, we aim in this paper to study the
circumstellar matter (CSM) that is expected to be blown during
the merger process in the DD scenario.
To distinguish CSM that might have been expelled prior to the merger
process from the material expelled from the accretion disk
formed during the merger process, we term the latter disk-originated
matter (DOM).
In this paper we study the DD scenario, postponing the CD scenario
to a future paper.

The paper is organized as follows. In section \ref{sec:previous} we list some previous studies
of the DD scenario where a CSM or an extended envelope are formed.
In section \ref{sec:DOM} we concentrate on the formation of the
DOM, and in section \ref{sec:explosion} we study the implications
of an explosion inside the DOM and compare the different cases
with observations. Discussion and short summary are in section
\ref{sec:discussion}.

\section{Previous Studies of Pre-Explosion CSM}
\label{sec:previous}

Numerous studies (e.g. \citealt{Yoon2007,LorenAguilar2009,Dan2011,
Pakmor2012b,Raskin2012,Zhu2013}) have used smoothed particle
hydrodynamics (SPH) simulations to evolve a binary WD system from
first contact until the complete destruction of the donor.
This is referred to as the dynamical phase of the merger.
Some of these simulations are summarized in Table \ref{table:models comparison}.
While different initial conditions are used in these simulations,
most results are common to all them.
After a time $t_{\rm dyn} \sim 100\s$ that corresponds to several
orbital periods, the donor is completely disrupted.
The merged product consists of the more massive primary WD, which
remains almost intact, with the material of the donor residing in a hot
corona surrounding it and in a nearly Keplerian disk extending out to $\sim 0.1\Rsun$.
In addition, a tidal tail with a mass of $\sim 10^{-3}\Msun$ is
unbound from the system (e.g. \citealt{RaskinKasen2013}).
A different possibility is the violent merger scenario of \cite{Pakmor2012b}.
They found that in some dynamical mergers hot spots are formed,
which experience a thermonuclear runaway.
The resulting explosion is highly asymmetrical \cite{Pakmor2012b}.
However, such asymmetrical explosions are in contradiction with
\revision{some} resolved SNR in the Galaxy and the Magellanic
clouds (\citealt{Lopezetal2011};
\revision{see more in section \ref{sec:discussion}).}
\cite{Moll+2014} found that the variations in light-curves and spectra
with viewing angle of explosions triggered by a violent merger are larger
than observed variations.

\begin{table*}

\caption{Comparison of outcomes of the DD scenario}
\label{table:models comparison}

\begin{tabular}{l|lllll}

\hline

{} & 
{A} & 
{B} & 
{C} & 
{D} & 
{E} \\

\hline

{Papers} &
{\cite{Shen2012};} &
{\cite{vanKerkwijk2010};} &
{\cite{Ji2013};} &
{\cite{Pakmor2012b}} &
{This paper} \\ 

{} &
{\cite{Schwab2012}} &
{\cite{Zhu2013}} &
{\cite{Beloborodov2013}} &
{} &
{} \\

\hline

{Fiducial} &
{$0.9+0.6$} &
{$0.8+0.4$ or $0.6+0.6$} &
{$0.6+0.6$} &
{$1.2+0.9$} &
{None} \\

{system(s) ($\rm M_{\sun}$)} &
{} &
{} &
{} &
{} &
{} \\

{Assumptions} &
{Axisymmetry;} &
{Spherical; wind carries} &
{Axisymmetry;} &
{Off-center hotspots} &
{AM carried by jets or disk-} \\

{} &
{$\alpha$-viscosity} &
{all AM with $E=0$} &
{MRI viscosity} &
{detonate during merger} &
{wind with a terminal velocity} \\

{} &
{} &
{} &
{} &
{} &
{of $v_{\rm esc} \left( R_{\rm rem} \right)$} \\

{$t_{\rm acc}$ ($\rm s$)} &
{$3 \times 10^4$} &
{N/A} &
{$2 \times 10^4$} &
{N/A} &
{$\sim 10^4$} \\

\hline
\multicolumn{6}{l}{Remnant star$^{\rm a}$} \\

{$M_{\rm rem}$ ($\rm M_{\sun}$)} &
{$1.04$} &
{$0.96$ or $0.91$} &
{$1.12$} &
{$2$} &
{$\sim 1-1.5$} \\

{$R_{\rm rem}$ ($\rm R_{\sun}$)} &
{$0.007$} &
{$0.01$ or $0.02$} &
{$0.03$} &
{$0.03$ $^{\rm b}$} &
{$\la 0.02$} \\

{Explosion} &
{No explosion. Steady} &
{Explosion in similar-mass} &
{Unclear} &
{Explosion during} &
{Explosion is assumed to} \\

{} &
{carbon burning.} &
{systems, no explosion in} &
{} &
{dynamical merger} &
{check consequences} \\

{} &
{} &
{nonsimilar-mass systems} &
{} &
{} &
{} \\

\hline
\multicolumn{6}{l}{Expanded envelope$^{\rm c}$} \\

{$M_{\rm env}$ ($\rm M_{\sun}$)} &
{$0.46$} &
{N/A} &
{$0.06$} &
{N/A} &
{N/A} \\

{$R_{\rm env}$ ($\rm R_{\sun}$)} &
{$1.4$} &
{N/A} &
{$0.14$} &
{N/A} &
{N/A} \\

\hline
\multicolumn{6}{l}{Disk-Wind (DOM)$^{\rm d}$} \\

{$M_{\rm DOM}$ ($\rm M_{\sun}$)} &
{$6 \times 10^{-6}$} &
{$0.25$ or $0.29$} &
{$10^{-3}$} &
{$5 \times 10^{-3}$} &
{$0.04$} \\

{$v_{\rm DOM}$ ($\rm km \; s^{-1}$)} &
{N/A} &
{$\sim 2000$ $^{\rm e}$} &
{$\sim 3000$} &
{$\sim 2000$} &
{$\sim 5000$ $^{\rm f}$} \\

\hline

{Comments} &
{} &
{Optically thick DOM at} &
{Jets were proposed for} &
{Highly asymmetrical} &
{Such an outflow} \\

{} &
{} &
{$\sim 1-10 \Rsun$} &
{Kepler SNR} &
{explosion contradicts} &
{contradicts SN~2011fe} \\

{} &
{} &
{contradicts SN~2011fe} &
{\citep{TsebrenkoSoker2013}} &
{resolved Type Ia SNR} &
{(see Section \ref{sec:explosion})} \\

{} &
{} &
{\citep{Nugent+2011}} &
{DOM is expected} &
{\citep{Lopezetal2011}} &
{} \\

\end{tabular}

\flushleft
\textit{Notes.} AM is angular momentum, $v_{\rm esc}$ is the escape velocity from the remnant during the accretion
phase $v_{{\rm esc}}=\sqrt{2GM_{{\rm rem}}/R_{\rm rem}}$. \newline
$^{\rm a}${The primary with the accreted mass from the disk (not the SNR).} \newline
$^{\rm b}${\revision{At the time of explosion $t=610\s$ in \cite{Pakmor2012b}
				a dense gas from the destroyed WD extends up to distances of $\sim 0.015-0.03 \Rsun$,
				and more rarefied gas (seen at $t=612 \s$ in their figure 1) extends up to $\sim 0.03-0.045 \Rsun$.
				We therefore mark the typical radius of the remnant in their simulation as $0.03 \Rsun$.}} \newline
$^{\rm c}${Material surrounding the accreting WD.} \newline
$^{\rm d}${Material blown by the accretion disk, termed here DOM for disk-originated
				matter. In all systems an additional $\sim 10^{-3} \Msun$ is unbounded during
				the dynamical merger phase in the tidal tail.} \newline
$^{\rm e}${Material is ejected with zero total energy.
				$v_{\rm DOM} = v_{\rm esc} \left( R_{\rm DOM} \right)$} \newline
$^{\rm f}${Material is ejected with a terminal velocity of
				$v_{\rm DOM} = v_{\rm esc} \left( R_{\rm rem} \right)$} \newline

\end{table*}

An important property of systems following a dynamical merger
is a clear hierarchy of timescales.
The dynamical timescale of the merger $t_{\rm dyn} \sim \Omega^{-1}$
is of the order of seconds.
Next in order of magnitude is the viscous timescale of the
accretion disk, which characterizes the transport of disk mass
inwards and angular momentum outwards.
Modelling of this ``viscous phase'' in the merger is usually done using the
Shakura-Sunyaev $\alpha$-prescription \citep{ShakuraSunyaev1973},
giving a timescale of
$t_{\rm visc} \sim \alpha^{-1} \left( R_{\rm disk} / H \right) t_{\rm dyn}$
with $R_{\rm disk}$, $H$ and $t_{\rm dyn}$ the disk radius, scale
height and dynamical timescale, respectively (e.g. \citealt{vanKerkwijk2010}).
Taking a suitable value of $\alpha = 0.01-0.1$ gives a viscous
timescale of $t_{\rm visc} \sim 10^3-10^4 \s$.
Largest in the timescale hierarchy is the thermal timescale of the
merger product, which can vary depending on the possible existence
of stable carbon burning but is always of the order of years or above.
This timescale hierarchy $t_{\rm dyn} \ll t_{\rm visc} \ll t_{\rm th}$
justifies the breakdown of studies into the different merger phases,
taking different assumptions for each.
We focus on the implications of an explosion occurring during or after the viscous phase.
Our results do not apply to models where the delay between the merger and the
explosion is very long, up to millions of years, as expected if rotation
keeps the remnant stable even when it exceeds $M_{\rm Ch}$
(e.g.,\citealt{TornambePiersanti2013}).

\cite{Schwab2012} and \cite{Zhu2013} investigated the evolution of
the merger product following the dynamical phase.
Of the various systems studied by \cite{Schwab2012}, none developed
a thermonuclear runaway.
\cite{Zhu2013} evolved the remnant of a $0.8\Msun+0.4\Msun$ fiducial
system, and found that a mass of $0.25\Msun$ was expelled -- a value
much higher than for similar systems in \cite{Schwab2012}.
If the assumptions of \cite{Zhu2013} are accepted, the removed mass
is the matter that is not accreted due to angular momentum conservation.
This matter is marginally unbound, and can correspond to the giant
structure of \cite{Schwab2012}, which is not unbound but is
significantly extended around the degenerate merger.

Severe constraints have already been placed on the DD scenario as a
frequent channel for typical SNIa explosions by the above mentioned
studies, as well as by other studies.
The companion papers of \cite{Moll+2014} and \cite{Raskinetal2013}
stress the large variations in viewing angle for an explosion during
or directly after the dynamical merger, respectively.
\cite{Raskinetal2013} also show that an explosion inside the accretion
disk leads to excessive $^{56}$Ni production and peculiar observables
for merger products with a total mass exceeding the Chandrasekhar mass,
while \cite{Schwab2012} show that less massive merger products fail to
explode.
The failure to detonate could mean that the remnant eventually collapses to a
neutron star \citep{SaioNomoto1985}.
However, since other studies do support the possibility of the DD
scenario leading to standard SNIa explosions (e.g., \citealt{vanKerkwijk2010,Zhu2013}),
it seems plausible to further address the implications of matter
launched by winds or jets during the viscous phase, on an
explosion at a somewhat later time, when the disk matter has been
accreted or expelled.
Since it is focused on this later time period, our study can be viewed
as complementary to studies of the observational properties of an
explosion during or directly after the dynamical merger \citep{Moll+2014,Raskinetal2013}.

In the studies mentioned above, a common feature is the existence
of substantial disk-originated matter (DOM) in the vicinity of
the exploding object.
The collision of exploded ejecta with the DOM will influence the
inferred size of the exploding object, making it $\ga 1\Rsun$.
For the fiducial model of \cite{Schwab2012}, the DOM is opaque up
to $>1\Rsun$ (e.g. last panel of their Fig. 5).
For the models of \cite{Zhu2013} the same holds, as will be shown
in the next section.
This is a problematic result for regular SN~Ia, whose size is an order
of magnitude smaller, e.g. SN~2011fe \citep{Nugent+2011,
Bloom+2012,PiroNakar2014}.
We now turn to study the observational signatures of this DOM.

\section{Removal of Disk Material}
\label{sec:DOM}

We consider here a model with the assumption, based on stellar winds
and velocities of jets from accretion disks, that the terminal velocity
of the wind equals the escape velocity from the vicinity of the more massive WD
\begin{equation}
	v_{\rm DOM} \simeq
	v_{\rm terminal} \simeq v_{\rm esc} \left( R_{\rm rem} \right) \simeq 5000 \kms,
\label{eq:v_terminal}
\end{equation}
where $R_{\rm rem}$ is the radius of the remnant during the accretion phase.
We assume spherical symmetry as well, despite the expectation for a
bipolar structure due to jets and/or disk winds.
The radius of the DOM is
\begin{equation}
R_{\rm DOM} \simeq
v_{\rm DOM} t =
72 \left(\frac{v_{\rm DOM}}{5000 \kms}\right)
  \left(\frac{t}{10^4 \s}\right) \Rsun.
\label{eq:DOM radius}
\end{equation}
Mass is expelled at a range of velocities, so it will be useful to
define the average DOM density
\begin{align}
\bar{\rho}_{{\rm DOM}} \simeq &
1.5 \times 10^{-7} \left(\frac{M_{{\rm DOM}}}{0.04\Msun}\right) \times \nonumber \\
  & \left(\frac{v_{\rm DOM}}{5000 \kms}\right)^{-3}
  \left(\frac{t}{10^4 \s}\right)^{-3} \g \cc,
\label{eq:DOM density}
\end{align}
where $M_{{\rm DOM}}$ is the expelled mass.
The value taken for this mass is based on the typical ratio of mass outflow
rate in jets to accretion rate in systems observed to launch jets,
$\sim 5-10 \%$, and on a typical WD companion mass of $0.4-0.8 \Msun$.
Taking free electron scattering opacity
($\kappa_{{\rm T}}=0.2~\rm{cm^{2}\;g^{-1}}$) as a lower bound, the optical
depth is
\begin{align}
\tau \ga
\kappa_{{\rm T}}\bar{\rho}_{{\rm DOM}}r_{{\rm DOM}} =
1.5 \times 10^{5} \left(\frac{M_{{\rm DOM}}}{0.04\Msun}\right) \times \nonumber \\
  \left(\frac{v_{\rm DOM}}{5000 \kms}\right)^{-2}
  \left(\frac{t}{10^4 \s}\right)^{-2}
  \left( \frac{\kappa}{\kappa_{\rm T}} \right).
\label{eq:DOM optical depth}
\end{align}
This shows that the wind-blown DOM is opaque throughout the viscous phase.
If an explosion takes place inside this wind, it will be observed as an
explosion of an object of size $\ga 1 \Rsun$, in contradiction with
SN~2011fe whose progenitor radius is limited to $R \la 0.1 \Rsun$
\citep{Bloom+2012,PiroNakar2014}.
This will be elaborated upon in the next section.
Note that the wind-blown DOM is opaque even if a much smaller mass
($\sim10^{-3}\Msun$) is blown, such as through tidal tail formation
\citep{RaskinKasen2013} or magnetized outflow \citep{Ji2013,Beloborodov2013},
though for these examples the mass loss is far from spherical.

\cite{Zhu2013} use different assumptions on the expelled material during the viscous phase.
The DOM is assumed to leave with zero total energy and so has an outflow
velocity of $v_{{\rm esc}}=\sqrt{2GM_{{\rm rem}}/R_{{\rm DOM}}}$, where
$M_{{\rm rem}}$ is the remnant mass.
The mass of the DOM is also larger, $\sim 0.2 \Msun$.
This leads to a more compact DOM, of radius
\begin{align}
R_{\rm DOM} & \simeq
\left(\frac{3}{2} \sqrt{2GM_{\rm rem}}\cdot t + R_{0}^{\frac{3}{2}}\right)^{\frac{2}{3}} \nonumber \\
& \sim 5.6 \left(\frac{M_{{\rm rem}}}{1 \Msun}\right)^{\frac{1}{3}}
  \left(\frac{t}{2 \times 10^4 \s}\right)^{\frac{2}{3}} \Rsun,
\label{eq:DOM radius for van Kerkwijk}
\end{align}
so the density of the DOM is higher and it is likewise opaque throughout the viscous phase.

\section{Explosion Inside Disk-Originated Matter (DOM)}
\label{sec:explosion}

If the merger remnant would explode during the viscous phase of the
accretion disk as in the scenario proposed by \cite{vanKerkwijk2010},
the high-velocity exploded material (ejecta) will shock the DOM and
generate an observable signal. We discuss here two effects of the
shocked DOM on observations.
(1) The extra thermal energy in the shocked DOM will lead to a larger
inferred progenitor radius.
(2) The passage of the shock wave through the DOM will generate a
transient signal.

\subsection{Inferred Progenitor Radius}
\label{subsec:progenitor radius}

The fresh SN ejecta is radiation-dominated, and adiabatic expansion
reduces thermal energy as $1/r$. By $1 \Rsun$ the thermal energy is
reduced to $\sim 0.02$ times its initial value for an initial WD
radius of $R_{\rm WD} = 0.02 \Rsun$.
In our model the DOM, with a mass of $M_{\rm DOM} \sim 0.04 \Msun$,
is shocked at $R_{\rm DOM} \sim 10-100 \Rsun$.
The relative velocity between the DOM, with $v_{\rm DOM} \la 5000 \kms$,
and the ejecta, with $v_{\rm ej} \sim 15,000 \kms$, implies that the
amount of kinetic energy that is transferred to thermal energy during
the collision is
$E_{\rm shock} \simeq 5\times 10^{50} (M_{\rm DOM} / M_{\sun}) \erg$.
This can be much larger than the thermal energy of the ejecta just
before it hits the DOM,
$\simeq 5\times 10^{50} (R_{\rm prog}/R_{\rm DOM}) \erg$, where
$R_{\rm prog}$ is the progenitor's radius.
We took the initial thermal energy of the exploding WD to be half the
explosion energy (the rest is kinetic energy).

When the thermal energy content of the gas is used to infer the initial
radius at later times, by using the radiation before $^{56}$Ni decay
becomes dominant (e.g., \citealt{Nugent+2011}), the observations will
be interpreted as if the progenitor radius was
\begin{align}
R_{\rm prog,i} & \ga
R_{\rm DOM} \frac{M_{\rm DOM}}{M_{\rm WD}} \nonumber \\
& = 0.3 \left( \frac{R_{\rm DOM}}{10 \Rsun} \right)
  \left( \frac{M_{\rm DOM}}{0.04 \Msun} \right) 
  \left( \frac{M_{\rm WD}}{1.4 \Msun} \right)^{-1} \Rsun.
\label{eq:fake r_prog}
\end{align}
Such a large radius is ruled out for SN~2011fe (e.g. \citealt{Nugent+2011,
Bloom+2012,PiroNakar2014}).
The constraints on SN~2011fe of $R_{\rm WD} \la 0.1 \Rsun$ limits
the DOM mass at $R_{\rm DOM} \sim 100 \Rsun$ to $\la 10^{-3} \Msun$.
Note that using the more massive and compact DOM derived from the
\cite{Zhu2013} fiducial model gives an inferred progenitor radius,
$R_{\rm prog,i} \simeq 0.7 \Rsun$, which also exceeds the constraint on SN~2011fe.

The derivation of equation (\ref{eq:fake r_prog}) assumes that the thermal
energy due to the ejecta-DOM collision is distributed in the entire ejecta,
as the thermal energy in the explosion itself.
However, the thermal energy of the collision is distributed in the DOM and
the outer part of the ejecta.
The radiation that diffuses out comes from these outer parts.
This implies that thermal energy that is radiated at early hours is much
larger than if the thermal energy would have been distributed in the entire ejecta.
Consequently, the value given in equation (\ref{eq:fake r_prog}) is a lower
bound on the inferred  progenitor radius at early hours.

\subsection{Transient UV Signal}
\label{subsec:signal}

The front of the SN ejecta moves at a velocity of $\sim 20,000 \kms$.
A typical value for the velocity of the ejecta shocking the DOM is found by
taking an exponential density profile for the ejecta \citep{DwarkadasChevalier1998}
and searching for the velocity above which the total mass of the ejecta
is about $M_{\rm DOM} \sim 0.04 \Msun$.
This gives $v_{\rm ej} \simeq 15,000 \kms$.
The shock passes through the DOM in a dynamical time
\begin{equation}
t_{\rm dyn} \simeq 5 \times 10^3
\left( \frac{v_{\rm ej} - v_{\rm DOM}}{2 v_{\rm DOM}} \right)^{-1}
\left( \frac{\Delta t_{\rm exp}}{10^4 \s} \right) \s,
\label{eq:shock t_dyn}
\end{equation}
where $\Delta t_{\rm exp}$ is the time between the beginning of DOM formation
in the viscous phase and the explosion.

After ejecta-DOM collision, expansion continues, and most of the thermal
energy is lost to adiabatic expansion of the DOM and ejecta, now a combined medium.
Mostly photons which can diffuse on a time shorter than about $t_{\rm dyn}$ will escape.
More photons will diffuse later, and lead to the inferred large progenitor radius
as discussed in section \ref{subsec:progenitor radius}.
The diffusion time is given by
\begin{equation}
t_{\rm dif} = \tau \frac{l_{\rm dif}}{3c} =
\frac{l_{\rm dif}^2 \kappa \rho_{\rm s}}{3c}
\label{eq:diffusion time}
\end{equation}
(e.g. \citealt{Kasen2010}), where $\rho_{\rm s}$ is the shocked DOM density and
$l_{\rm dif}$ is the layer through which photons diffuse.
For a radiation pressure dominated gas, $\rho_{\rm s} = 7\rho_{\rm DOM}$.
Equating $t_{\rm dif} = t_{\rm dyn}$ and using
$\rho_{\rm DOM} \approx \bar{\rho}_{\rm DOM}$ gives a diffusion distance of
\begin{align}
l_{\rm dif} \sim 1.2
\left( \frac{M_{\rm DOM}}{0.04\Msun} \right)^{-\frac{1}{2}}
\left( \frac{\Delta t_{\rm exp}}{10^4~{\rm s}}\right)^{2}
\left( \frac{v_{\rm ej}}{15000 \kms} \right)^{\frac{3}{2}} \times \nonumber \\
\left( \frac{v_{\rm ej} - v_{\rm DOM}}{2 v_{\rm DOM}} \right)^{-2}
\left( \frac{\kappa}{\kappa_{\rm T}} \right)^{-\frac{1}{2}}
\Rsun.
\label{eq:diffusion length}
\end{align}
The diffusion distance cannot be larger than the length of the shocked DOM,
which is approximately $l_{\rm DOM,shocked} \sim 0.1 R_{\rm DOM}$.
The last equation and the following ones assume that indeed
$l_{\rm dif} \leq l_{\rm DOM,shocked}$, which is valid therefore for
$\Delta t_{\rm exp} \la 10^5 \s$.
For a larger $\Delta t_{\rm exp}$, instead of equation (\ref{eq:diffusion length})
we have $l_{\rm dif} \simeq l_{\rm DOM,shocked}$.
The energy density in the diffusion volume is $\epsilon_{\rm s} = 3p_{\rm s}$.
Since the density of the unshocked DOM is much less than that of the ejecta,
the pressure of the shocked gas is roughly the ram pressure,
\begin{equation}
p_{\rm s} \approx
\frac{6}{7} \rho_{\rm DOM} \left( v_{\rm ej} - v_{\rm DOM} \right)^2.
\label{eq:shock pressure}
\end{equation}
The diffusion volume is a shell of thickness $l_{\rm dif}$ at a radius of $r_{\rm DOM}$.
The radiation energy diffusing outwards from this volume is
\begin{align}
E \approx & 0.5 \cdot 0.5 \cdot 3 p_{\rm s} \cdot 4\pi R_{\rm DOM}^2 l_{\rm dif} \nonumber \\
\simeq & 1.7 \times 10^{48}
\left( \frac{M_{\rm DOM}}{0.04\Msun} \right)^{\frac{1}{2}}
\left( \frac{\Delta t_{\rm exp}}{10^4~{\rm s}}\right)
\left( \frac{v_{\rm ej}}{15000 \kms} \right)^{\frac{1}{2}} \times \nonumber \\
& \left( \frac{v_{\rm DOM}}{5000 \kms} \right)
\left( \frac{v_{\rm ej} - v_{\rm DOM}}{10000 \kms} \right)
\left( \frac{\kappa}{\kappa_{\rm T}} \right)^{-\frac{1}{2}} \erg,
\label{eq:shock energy}
\end{align}
where a factor of $0.5$ was taken to account for half of the photons diffusing
inwards, and another factor of $0.5$ was taken to account for roughly half of
the energy being diffused, with the remainder going to adiabatic expansion.
The average luminosity during $t_{\rm dyn}$ is
\begin{align}
L \approx \frac{E}{t_{\rm dyn}}
\simeq 3 \times 10^{44}
\left( \frac{M_{\rm DOM}}{0.04\Msun} \right)^{\frac{1}{2}}
\left( \frac{v_{\rm ej}}{15000 \kms} \right)^{\frac{1}{2}} \times \nonumber \\
\left( \frac{v_{\rm ej} - v_{\rm DOM}}{10000 \kms} \right)^{2}
\left( \frac{\kappa}{\kappa_{\rm T}} \right)^{-\frac{1}{2}}
\ergs,
\label{eq:shock luminosity}
\end{align}
which is independent of the delay time between DOM formation and explosion, under
the simplifying assumptions taken here, as long as $\Delta t_{\rm exp} < 10^5 \s$.
This gives a luminosity of $L \sim 10^{44} \ergs$ from the collision of SN ejecta
with the material expelled during the viscous phase.
We note that since the value $v_{\rm ej} = 15000 \kms$ was chosen as a minimal
velocity for the outer part of the ejecta which shocks the DOM, the luminosity
obtained by using a more detailed ejecta profile might be larger.
The effective temperature is
\begin{align}
T_{\rm eff} \approx 3 \times 10^{5}
\left( \frac{M_{\rm DOM}}{0.04\Msun} \right)^{\frac{1}{8}}
\left( \frac{\Delta t_{\rm exp}}{10^4~{\rm s}}\right)^{-\frac{1}{2}} \times \nonumber \\
\left( \frac{v_{\rm ej}}{15000 \kms} \right)^{-\frac{3}{8}}
\left( \frac{v_{\rm DOM}}{5000 \kms} \right)^{\frac{1}{2}}
\left( \frac{\kappa}{\kappa_{\rm T}} \right)^{-\frac{1}{8}} \K,
\label{eq:shock T_eff}
\end{align}
which is a UV transient lasting for a few hours, well before the supernova's peak
luminosity at about 20 days.
Only $\sim 10^{-5}$ of the radiation is in the visible range, amounting to only
$L_{\rm V} \sim 2 \times 10^{40} \ergs$.

\revision{Using high cadence observations, \cite{Siverd+2014} put an upper limit on
any short-term luminosity variations of SN~2014J of
$L < 8.7 \times 10^{36} \ergs$ in the \textit{R}-band.
Using this in equation (\ref{eq:shock luminosity}) we can set a limit of
$M_{\rm DOM} \la 4 \times 10^{-6} \Msun$ for a velocity difference of only
$v_{\rm ej} - v_{\rm DOM} = 1000 \kms$.
This tight limit on DOM mass rules out an accretion event if the explosion
occurred more than few minutes after merger.
\cite{Goobar+2014} show that a model with a large progenitor radius
$\ga 1 \Rsun$ nevertheless fits these observations, when fixing the early
lightcurve time dependence according to the fireball model ($L \propto t^2$).
A progenitor of $\sim 0.02 \Rsun$ is also possible when relaxing this constraint,
however, and the validity of the $t^2$ time dependence as a general property
of early SN~Ia lightcurves is an open question (e.g., \citealt{PiroNakar2013,Firth+2014}).
If indeed the progenitor is large, our results indicate that the explosion must
take place within few hundreds seconds to comply with the luminosity limit of
\cite{Siverd+2014}.}

The derivation above assumed our wind-blown DOM model as summarized in column E
of Table \ref{table:models comparison}.
If instead we use the DOM scenario of \cite{Zhu2013} described in section
\ref{sec:previous} (column B of Table \ref{table:models comparison}), we find
that the smaller radius of this DOM leads to a shorter dynamical time, $\sim 400 \s$.
The higher density also means a smaller part of the shocked DOM contributes to
the diffused energy, so that the total diffused energy is smaller, $E \simeq 10^{48} \erg$.
The luminosity, however, is larger than our DOM model,
$L \simeq 2.5 \times 10^{45} \ergs$, because of the shorter dynamical time.
Since the radius of the shocked DOM is smaller the effective temperature is
higher and the diffusion of the shock energy is seen as an X-ray transient
lasting for several minutes rather than a longer UV transient as described above.

\section{Discussion and Summary}
\label{sec:discussion}

We have studied some implications of mass ejection during the merger process in
the double-degenerate (DD) scenario for SNe~Ia,
\revision{for cases where explosion occurs shortly, but not promptly,
after merger, i.e., in the time range $10^3 \s \la t_{\rm exp} \la 1~{\rm day}$
after merger.}
As the two WDs merge, the lighter one is destroyed and forms an accretion disk
around the more massive WD.
Angular momentum and energy must be removed from the merger remnant during the
accretion process of the merger.
The accretion disk exists for hours (its viscous time), and is expected to blow
a wind and/or launch jets that carry away energy and angular momentum.
In some models, the gas of the destroyed WD is instead inflated to a large
envelope around the massive WD.
Some possibilities for the evolution of the merger during the viscous phase are
listed in Table \ref{table:models comparison}.
Column E is our proposed outflow structure.
The matter that is expelled by the disk is termed disk-originated matter (DOM).

In section \ref{sec:DOM} we assumed that the DOM is expelled spherically.
This of course is not the case, but is adequate for our approximate derivation.
The typical size and density in the DOM under two sets of assumptions, summarized in columns
B and E of table \ref{table:models comparison}, are given in section \ref{sec:DOM}.

As the merger remnant explodes, the ejecta from the explosion shock the DOM,
and kinetic energy is transferred to thermal energy.
In section \ref{sec:explosion} we studied two consequences of this interaction
if explosion occurs within about a day from merger.
In section \ref{subsec:progenitor radius} we concluded that such an interaction
will lead to an inferred progenitor radius of $\ga 0.1 \Rsun$.
This is in contradiction with the smaller progenitor radius inferred for
SN~2011fe \citep{Nugent+2011,Bloom+2012,PiroNakar2014}.
In section \ref{subsec:signal} we study the transient signal emerging from this
interaction, and find it to be a UV transient lasting up to a few hours.
\revision{We then used the derived luminosity to argue that SN~2014J could not
have exploded in the DD scenario if the explosion occurred during the viscous phase.
The time of explosion in a DD model for SN~2014J can be further constrained,
up to tens of years after merger (see below).}

If the explosion occurs before the complete destruction of the
lighter WD, our calculations are not applicable.
In that case the explosion will be highly non-spherical
\citep{Pakmor2012b,Moll+2014,Raskinetal2013}, which contradicts
the morphology of \revision{some} close (Galactic and in
the Magellanic Clouds) young SN remnants.
\revision{We can list several SNR that show no dipole asymmetry
(images from the Chandra SNR catalogue
\footnote{\href{http://hea-www.cfa.harvard.edu/ChandraSNR/}{http://hea-www.cfa.harvard.edu/ChandraSNR/}};
\citealt{Seward2004}).
SNRs that have (almost) spherical morphology include the Tycho SNR
and SNR 0509-67.5.
The simulations of \cite{Papishetal2015} suggest that the presence of any
companion will lead to large dipole asymmetry, not observed in these SNRs.
The SNRs G1.9+0.3 and Kepler have two opposite `Ears', but don't show a
prominent dipole morphology.
These also seem not to result from explosions taking place before the
complete destruction of the companion WD. 
One SNR that does show a dipole asymmetry is SN~1006. However,
\cite{Papishetal2015} find this dipole structure to be different from the
one expected due to the effect of the He~WD companion on the ejecta in the DDet scenario.
Further simulations are needed to follow the SNRs that result from
explosions taking place before complete donor destruction in the DD scenario.
Some SNRs are old and have no well-defined shapes, while in some other
SNRs the dipole asymmetry might result from an interaction with the ISM;
we do not refer to them here.}

The implications at different times of explosion since merger are
summarized in Table \ref{table:Dt_exp vs observations}.
This study focused on an explosion during the
viscous timescale of the accretion disk around the massive WD,
$t_{\rm visc} \approx 10^3 - 10^5 \s$.
Our study is complementary to those of \cite{Moll+2014} and
\cite{Raskinetal2013} who explored the consequences of an explosion
during or immediately after the dynamical merger.
Put together, the regime they explored and the regime we do
in the present study leave a small window for the DD scenario
to be compatible with observations of typical SN~Ia, unless a much larger
time delay until an eventual explosion is considered.
If the time delay to explosion is crudely $1 - 20 \days$,
we expect a strong optical extra peak before the
maximum luminosity, when the ejecta catch up to the DOM. If the
delay is greater, up to tens of years, a late peak might occur.
Delays of over tens of years might not have a prominent
observational signature from collision with the DOM, but require a
delay mechanism, such as rotation (e.g.,
\citealt{TornambePiersanti2013}). Such a long delay will allow a
merger remnant of near-Chandrasekhar mass to develop. An exploding
near-Chandrasekhar mass WD is compatible with recent findings that a large
fraction of SN~Ia masses are peaked around $1.4 \Msun$ \citep{Scalzoetal2014},
as required also for manganese nucleosynthesis \citep{Seitenzahletal2013}.
\begin{table*}

\caption{Observational effects of the DD scenario with different explosion times
\label{table:Dt_exp vs observations}}

\begin{tabular}{lll}

{$\Delta t_{\rm exp}$} &
{Outcome} &
{Observations} \\

\hline

{$< t_{\rm visc}$} &
{Asymmetrical explosion} &
{Symmetric SNRs} \\

{$\sim t_{\rm visc} \la 1 ~{\rm day}$} &
{Inferred $R_{\rm exp} > 0.1 \Rsun$} &
{SN~2011fe with $R_{\rm exp} \la 0.1 \Rsun$} \\

{$\sim 1 - 20 \days$} &
{Extra radiation, possible peak before maximum} &
{Not observed} \\

{$\sim 20 \days - 10\yr$} &
{Late peak} &
{Not observed} \\

{$\gg 10\yr$} &
{No effect, but this requires an explosion delay mechanism} &
{-} \\

\end{tabular}

\medskip
\flushleft
$t_{\rm visc} \sim 10^3-10^4 \s $ is the viscous time. \newline
$t_{\rm max} \simeq 20 \days$ is the time to maximum light.

\end{table*}

To summarize, our results put a stringent constraint on the time of explosion
\revision{of any DD scenario where a violent merger does not occur.
For systems where the conditions for explosion are not reached during the
dynamical merger, the subsequent creation of an extended structure of DOM via
jets or a disk-wind is unavoidable on account of angular momentum conservation.}
If the explosion is to occur at this stage, the additional radiation from
cooling of the \revision{gas heated in the ejecta-DOM shock} will be
observable, and inflate the value of the inferred progenitor radius to $\ga 1 \Rsun$.
These considerations suggest that possible explosion times after the dynamical
merger are at least one day after merger, when the temperature of the accreted gas
has dropped beneath the requirements for carbon-ignition.
\revision{Non-violent DD models which propose an explosion of the merger remnant should
therefore contain a suitable delay mechanism.}

\section*{ACKNOWLEDGEMENTS}
\label{sec:acknowledgements}

\revision{We thank our anonymous referee for valuable comments.}
This research was supported by the Asher Fund for Space Research at the Technion, and the US-Israel Binational Science Foundation.
This work was also partially supported by MCINN grant AYA2011--23102, and by the European Union FEDER funds.

{}

\end{document}